\documentstyle[12pt,a4wide]{article}
\begin{document}

\thispagestyle{empty}
\begin{flushright}
MZ-TH/97-16\\
hep-ph/9704396\\
April 1997
\end{flushright}
\vspace{0.5cm}
\begin{center}
{\Large\bf Perturbative relations between $e^+e^-$ annihilation and}\\[.3truecm]
{\Large\bf $\tau$ decay observables including resummation effects}\\[1truecm]
{\large S.~Groote,$^1$ J.G.~K\"orner$^1$ and A.A.~Pivovarov$^{1,2}$}\\[.7cm]
$^1$ Institut f\"ur Physik, Johannes-Gutenberg-Universit\"at,\\[.2truecm]
  Staudinger Weg 7, D-55099 Mainz, Germany\\[.5truecm]
$^2$ Institute for Nuclear Research of the\\[.2truecm]
  Russian Academy of Sciences, Moscow 117312
\vspace{1truecm}
\end{center}

\begin{abstract}
By exploiting the analyticity properties of the two-point current-current 
correlator we obtain numerical predictions for the $e^+e^-$ moments in 
terms of the $\tau$ decay rate. We perform a partial resummation of the 
pertinent perturbative series expansion by solving the renormalization 
group equation for Adler's function. Our predictions are renormalization 
scheme independent but depend on the order of the perturbative 
$\beta$-function expansion. The analysis involves the unknown five-loop 
coefficient $k_3$ for which we give some new estimates.
\end{abstract}

\newpage\noindent
The interpretation of experimental strong interaction data in terms of QCD 
and the Standard Model requires higher and higher accuracy in the 
theoretical input~\cite{altrev,Shifman}. There is a need to improve 
on the precision of the predictions of perturbative QCD by developing a 
framework that allows one to go beyond finite orders of perturbation theory.
This is indispensable if one wants to make progress in precision fits to 
data of existing and especially future experiments. New high order results 
in fixed order perturbation theory are becoming more difficult to 
come by because of overwhelming technical difficulties.\footnote{One of the 
recent achievements was the calculation of the four-loop 
$\beta$-function~\cite{Ritbergen} and the four-loop anomalous mass 
dimension~\cite{mass4}.} What is needed is to develop techniques that allow 
one to go beyond the existing finite order perturbative results. Attempts in 
this direction include finite order predictions based on the Pad\'e 
approximation~\cite{Pade}, different optimizations of perturbation 
theory~\cite{KatSta}, and infinite resummation procedures based on 
particular properties of perturbation series like renormalon 
methods~\cite{AkhZak,AltNasRidRen}. Another approach consists in the use of 
the renormalization scheme freedom to parameterize infrared contributions 
to physical observables in the renormalon 
approximation~\cite{Khoze,KraPivRen,PenPivRen}. Latter approaches can serve 
as an alternative to the renormalon calculus. 

The analysis of the moments of the $e^+e^-$ rate and, in conjunction with 
it, the analysis of the $\tau$ decay rate has a long-standing history both 
in theory and experiment. The accuracy in the determination of the $\tau$ 
decay rate and its decay characteristics has been continuously improving 
while there is now hope that there will be more precise data on the 
$e^+e^-$ rate in the low energy domain in the near future~\cite{aleph}. 
Because of the availability of a large number of terms in the perturbative 
QCD expansion and the simplicity of the analyticity structure of the 
underlying Green's function, the analysis of the $e^+e^-$ annihilation 
process has advanced to a highly sophisticated stage. The analyticity 
structure is simple because the process is related to the two-point 
correlator of gauge invariant current correlators, the analyticity 
structure of which is determined by the K\"all\'en--Lehmann representation.

In a recent paper we advocated the idea to directly compare physical 
observables within fixed order perturbative QCD. Using this approach we 
determined moments of the $e^+e^-$ annihilation rate in terms of the $\tau$ 
decay rate~\cite{ptest}. This eliminates the problem of scheme dependence 
within finite order perturbative QCD (especially if the moments are taken at 
the same or comparable scale). In a second paper we exploited analyticity 
properties of the two-point current correlator to partially resum the 
perturbation series for a new analysis of $\alpha_s$~\cite{resum} (see 
also~\cite{tau}). In the present paper we combine the approaches 
of~\cite{ptest} and~\cite{resum} and present the results of an analysis 
which expresses the moments of $e^+e^-$ rate functions in terms of the 
$\tau$ decay rate including resummation effects. We also discuss some 
general features of the solution of the renormalization group equation in 
the complex plane and speculate on estimates of higher order coefficients 
of the perturbation series in a renormalization group invariant manner.

We closely follow the notation introduced in~\cite{resum}. Let us begin by 
defining moments of the $e^+e^-$ annihilation rate in terms of the spectral 
density $R(s)$ according to
\begin{equation}\label{momdef}
R_n(s_0)=(n+1)\int_0^{s_0}\frac{ds}{s_0}\left(\frac{s}{s_0}\right)^nR(s). 
\end{equation}
For convenience we have normalized the moments such that $R_n(s_0)=1$ for 
$R(s)=1$. We define reduced moment functions by factoring out moments of 
the partonic Born term contribution $R_n^0(s_0)$. The moments of the Born 
term contribution $R_n^0$ are given by the leading order in the strong 
interaction. Nonperturbative (power suppressed) corrections and corrections 
due to other interactions (i.e.\ electro-weak corrections in the case of 
the semileptonic $\tau$ decay) may be absorbed in $R_n^0$~\cite{Braaten}. 
After factorization of the Born term moments the reduced moment functions 
$r_n(s_0)$ are given by
\begin{equation}\label{rndef}
R_n(s_0)=R_n^0(s_0)\left(1+\frac49r_n(s_0)\right).
\end{equation}

Our aim is to establish relations between different sets of observables 
(and not relations between observables and powers of the strong coupling 
constant). Further the relations should be independent of the choice of a 
particular renormalization scheme. It is therefore convenient to deal with 
scheme independent quantities from the outset.

Similar to Eq.~(\ref{rndef}) we define the reduced Adler's function $d(Q^2)$
\begin{equation}
D(Q^2)=\frac1{4\pi^2}\left(1+\frac49d(Q^2)\right)
\end{equation}
which we take to be an effective coupling constant. In the Euclidean domain 
the reduced Adler's function $d(Q^2)$ has the expansion
\begin{equation}
d(Q^2)=a(Q^2)+k_1a^2(Q^2)+k_2a^3(Q^2)+k_3a^4(Q^2)+\ldots
\end{equation}
where $a=9\alpha_s/4\pi$.

The running of the effective coupling constant $d(Q^2)$ is determined by 
the renormalization group equation which we write as~\cite{Gru}
\begin{equation}\label{RGE}
Q^2\frac{d}{dQ^2}d(Q^2)=\beta(d(Q^2)),
\end{equation}
where the perturbative series expansion of the $\beta$-function reads
\begin{equation}\label{betafun}
\beta(d)=-d^2(1+\rho_1d+\rho_2d^2+\rho_3d^3+\ldots).
\end{equation}
Due to the fact that we have chosen the reduced Adler's function as our 
expansion parameter, the coefficients $\rho_i$ are renormalization scheme 
independent quantities (see also~\cite{Max}). Numerically they are given by 
($N_c=3$)
\begin{eqnarray}\label{rhonum}
\rho_1&=&\frac{64}{81}\approx 0.790\nonumber\\
\rho_2&=&\frac{16531}{2916}+\frac{728}{81}\zeta(3)
  -16\zeta(3)^2+\frac{200}{27}\zeta(5)\approx 1.035\nonumber\\
\rho_3&=&-\frac{37096148}{59049}+\frac{4820288}{6561}\zeta(3)
  +\frac{12352}{81}\zeta(3)^2-256\zeta(3)^3\nonumber\\&& 
  -\frac{59800}{243}\zeta(5)+\frac{1600}{9}\zeta(3)\zeta(5)+2k_3
  \approx 2k_3-2.97953 
\end{eqnarray}
where we have taken the numerical values for the known coefficients in the 
$\overline{\rm MS}$ scheme from~\cite{ChetKuhn} together with the recently 
calculated four-loop $\beta$-function coefficient given in~\cite{Ritbergen}.
It should be kept in mind though that the final values of the coefficients 
$\rho_i$ are scheme independent even if they have been calculated in a 
specific scheme. The five-loop contribution $k_3$ in the coefficient $\rho_3$ 
is not yet known. Later on we shall present some estimates for $k_3$. We have 
set $n_f=3$ in the present application. Here we are interested in moments 
of $e^+e^-$ annihilation spectral density at the scale $Q^2=m_\tau^2$. The 
extension of our approach to other scales is straightforward.

The reduced Adler's function $d(m_\tau^2)$ itself can be determined from 
experimental data in terms of the integral representation
\begin{equation}\label{dtau}
d(m_\tau^2)=m_\tau^2\int_0^\infty\frac{r(s)ds}{(s+m_\tau^2)^2} 
\end{equation}
where $r(s)$ is the reduced spectral density. Note that after factorization 
of the Born term contribution, $r(s)$ need not be positive definite anymore. 
One therefore may encounter dramatic cancellations in the evaluation of the 
integral in Eq.~(\ref{dtau}) at the cost of the precision with which 
$d(m_\tau^2)$ can be determined. This is the reason why we do not use the 
integral respresentation Eq.~(\ref{dtau}) for our numerical estimates of 
$d(m_\tau^2)$.

The standard technique of contour integration in the complex plane allows 
one to obtain closed formulas for the reduced moments $r_n$ in terms of 
$d(m_\tau^2)\equiv d_\tau$ and the coefficients $\rho_i$ appearing in the 
perturbative expansion of the $\beta$-function in Eq.~(\ref{betafun}). The 
moments are given by
\begin{equation}\label{momcircle}
r_n(m_\tau^2)=\frac{n+1}{2\pi i}\oint_{|x|=1}x^np(m_\tau^2x)dx,  
\end{equation}
where $p(z)$ is the reduced vacuum polarization function. It is related to 
the reduced Adler's function by
\begin{equation}\label{pdef}
d(Q^2)=-Q^2\frac{dp(-Q^2)}{dQ^2}.
\end{equation}
After integrating Eq.~(\ref{momcircle}) by parts using Eq.~(\ref{pdef}), 
the moments can be represented by
\begin{equation}\label{momsplit}
r_n(m_\tau^2)=r_{\rm circ}(m_\tau^2)+\Delta_n(m_\tau^2).
\end{equation}
The first term on the right hand side of Eq.~(\ref{momsplit}) represents 
the surface term contribution. It does not depend on $n$ and is, in fact, 
the renormalization group improved spectral density (the discontinuity 
across the cut at $s=m^2_\tau\pm i0$) which can be calculated from 
$d(Q^2)=d(m_\tau^2e^{i\phi})$ using
\begin{equation}\label{spectr}
r_{\rm circ}(m_\tau^2)=\frac1{2\pi}
  \int_{-\pi}^\pi d(m_\tau^2e^{i\phi})d\phi.
\end{equation}
The second term in the partial integration is given by
\begin{equation}\label{spectrn}
\Delta_n(m_\tau^2)=\frac{(-1)^n}{2\pi}\int_{-\pi}^\pi e^{i(n+1)\phi}
  d(m_\tau^2e^{i\phi})d\phi
\end{equation}
and depends on the order $n$ of the moment under consideration. Latter 
contributions tend to be numerically suppressed because of the oscillatory 
factor $e^{i(n+1)\phi}$. Considered as functions of the variable $n$, the 
quantities $\Delta_n$ are oscillating functions. At discrete values for $n$ 
they can give quite irregular contributions to the moments. Numerically 
these contributions are suppressed but are not negligible.
 
We mention that the reduced moments cannot be approximated with any 
precision by an asymptotic expansion in $d_\tau$. For small enough values 
of $d_\tau$ the corresponding series converge~\cite{resum}. However, if we 
take the numerical values for $d_\tau$ derived below, they lie outside the 
circle of convergence and so the terms in Eq.~(\ref{momsplit}) cannot be 
represented as series in $d_\tau$. That is why we prefer to work directly 
with the integral representation given by Eqs.~(\ref{spectr}) 
and~(\ref{spectrn}) and not with a series expansion.

To leading order in the $\beta$-function the reduced moments can be 
presented in explicit form. With $\beta(d)=-d^2$ one can integrate the 
renormalization group equation~(\ref{RGE}) in the complex $Q^2$-plane using 
the starting value $d_\tau$ and obtains 
$d(m_\tau^2e^{i\phi})=d_\tau(1+id_\tau\phi)^{-1}$. The moments can then be 
represented as
\begin{equation}\label{ro}
r_n(m_\tau^2)=\frac1\pi\arctan(\pi d_\tau)+\frac{(-1)^n}{2\pi}
  \int_{-\pi}^\pi\frac{e^{i(n+1)\phi}d\phi}{1+id_\tau\phi}.
\end{equation}

If more terms are included in the $\beta$-function expansion, we write the 
general result in symbolic form as
\begin{equation}\label{moments}
r_n(m_\tau^2)=r_n(d_\tau;\beta)=r_n(d_\tau;\rho_1,\rho_2,\rho_3,\ldots).
\end{equation}
When expanded for small $d_\tau$ and a given order of the $\beta$-function 
expansion, these moments reproduce the known results of fixed order 
perturbation theory. However, here we consider a different approach which 
is not based on a series expansion. The functions $r_n(m_\tau^2)$ can be 
computed from Eq.~(\ref{momcircle}) to any given order of the 
$\beta$-function expansion without ever having to invoke a series 
expansion. When one goes beyond the leading order approximation of the 
$\beta$-function, the renormalization group equation~(\ref{RGE}) can still 
be solved, but the integrations in Eqs.~(\ref{spectr}) and~(\ref{spectrn}) 
become unwieldy. In these cases we proceed in numerical fashion.

In order to make contact with the semileptonic $\tau$ decay rate we take a 
particular combination of moments, namely
\begin{equation}\label{rtau}
r_\tau=2r_0(m_\tau^2)-2r_2(m_\tau^2)+r_3(m_\tau^2).
\end{equation}
Using the partially integrated representation Eq.~(\ref{momsplit}), this 
results in
\begin{equation}\label{contact}
r_\tau(m_\tau^2)=r_{\rm circ}(m_\tau^2)+2\Delta_0(m_\tau^2)
  -2\Delta_2(m_\tau^2)+\Delta_3(m_\tau^2)
  =r_{\rm circ}(m_\tau^2)+\Delta_\tau(m_\tau^2).
\end{equation}
This formula holds for perturbative QCD in the case of massless quarks 
where axial-vector and vector contributions are equal.

As a next step we invert Eq.~(\ref{contact}) numerically to obtain 
numerical values for $d_\tau$ by solving the differential 
equation~(\ref{RGE}) with $d_\tau$ as starting value. This is done by a 
systematic trial and error procedure using as many iterations as necessary. 
The whole procedure is repeated for successive orders in the 
$\beta$-function expansion. Using the experimental value 
$r_\tau=0.487\pm 0.011$~\cite{Pich2}, we obtain
\begin{equation}
d_\tau^{(0)}=0.3431\pm 0.0069,\quad d_\tau^{(1)}=0.3466\pm 0.0081,\quad
d_\tau^{(2)}=0.3561\pm 0.0097
\end{equation}
for increasing orders in the $\beta$-function expansion, where the 
superscript labels the order of the expansion. Once these values are known, 
we can predict the contributions of $r_{\rm circ}$ and $\Delta_n$ to the 
different reduced moments $r_n$. Predictions for the different 
contributions using the leading, first and second order accuracy for the 
$\beta$-function are given in the first three columns of Table~\ref{tab1},
\begin{table}\begin{center}
\begin{tabular}{|r||l|l|l||l|}\hline
$i=$&0&1&2&3\\\hline
$d_\tau$&$0.3431$&$0.3466$&$0.3561$&$0.3667$\\\hline
$r_{\rm circ}(m_\tau^2)$&$0.2619$&$0.2396$&$0.2352$&$0.2354$\\\hline
$\Delta_0(m_\tau^2)$&$0.1255$&$0.1369$&$0.1411$&$0.1448$\\
$\Delta_1(m_\tau^2)$&$0.0290$&$0.0174$&$0.0108$&$0.0073$\\
$\Delta_2(m_\tau^2)$&$0.0204$&$0.0196$&$0.0210$&$0.0227$\\
$\Delta_3(m_\tau^2)$&$0.0148$&$0.0129$&$0.0118$&$0.0110$\\
$\Delta_4(m_\tau^2)$&$0.0116$&$0.0105$&$0.0103$&$0.0106$\\\hline
$\Delta_\tau(m_\tau^2)$&$0.2251$&$0.2474$&$0.2519$&$0.2552$\\\hline
\end{tabular}
\caption{\label{tab1}
  Moment contributions for increasing $\beta$-function accuracy $i$}
\end{center}\end{table}
whereas in Table~\ref{tab2} we present the results in terms of the full 
moments.
\begin{table}\begin{center}
\begin{tabular}{|r||l|l|l||l|}\hline
$i=$&$0$&$1$&$2$&$3$\\\hline
$d_\tau$&$0.3431$&$0.3466$&$0.3561$&$0.3667$\\\hline
$R_0(m_\tau^2)$&$2.3444$&$2.3347$&$2.3344$&$2.3380$\\
$R_1(m_\tau^2)$&$2.2586$&$2.2284$&$2.2187$&$2.2157$\\
$R_2(m_\tau^2)$&$2.2509$&$2.2304$&$2.2277$&$2.2295$\\
$R_3(m_\tau^2)$&$2.2459$&$2.2245$&$2.2195$&$2.2190$\\
$R_4(m_\tau^2)$&$2.2432$&$2.2223$&$2.2182$&$2.2187$\\\hline
\end{tabular}
\caption{\label{tab2}Full moments for increasing $\beta$-function 
  accuracy $i$}
\end{center}\end{table}

For the orders $n=0$, $1$ and $2$ of the moments the values in 
Table~\ref{tab2} agree with estimates given in~\cite{ptest} on the basis of 
fixed order expansion in perturbation theory. We cannot give a prediction 
for the moment $r_{-1/2}$ within the present approach because the weight 
function used for constructing moments has to be analytically continued and 
must be single-valued in the vicinity of the cut along the positive 
semi-axis. This is the only singularity in the complex plane allowed for 
the vacuum polarization two-point function as given by the spectral 
representation. Though $\sqrt z$ is an analytical function in the complex 
plane with a cut, its values on different sides of the cut have opposite 
signs. Because of this, one cannot extract the discontinuity of the vacuum 
polarization function across the cut. 

As a note we want to illustrate the advantage of the resummation technique 
in comparison with ordinary perturbation theory. This illustration is done 
by giving a very simple example concerning the really non-trivial question 
whether results can be obtained by directly keeping larger number of terms. 
Consider two observables given by perturbative series in the some given 
scheme, namely
\begin{equation}\label{example1}
f(a)=a(1-a+a^2-\ldots\ )=\frac{a}{1+a}
\end{equation}
and
\begin{equation}\label{example2}
g(a)=a(1-2a+4a^2-\ldots\ )=\frac{a}{1+2a}.
\end{equation}
The functions $f(a)$ and $g(a)$ can be seen to be related by 
\begin{equation}\label{convergent}
g(f)=\frac{f}{1+f}=f(1-f+f^2-\ldots\ ).
\end{equation}
If we fit the right hand side of Eq.~(\ref{example1}) to an experimental 
value of about $f=0.6$, we get $a=1.5$. But for this value of the coupling, 
the series in Eq.~(\ref{example1}) diverges. So we cannot get $a$ from it 
without a proper resummation procedure that in this case is trivially given 
by the appended exact formula. Consequently we cannot get a prediction for 
$g$ using the series in Eq.~(\ref{example2}) in terms of $a$. On the other 
hand, the direct relation in terms of the series in Eq.~(\ref{convergent}) 
converges perfectly and gives an unambiguous result for $g$ in terms of 
measured $f$. Of course, for such an improvement to occur one has to 
analyze in detail the underlying theory and the origin of the series. 
Within our resummation procedure we are able to do so though we still
could not answer whether one can reexpand moments through $r_\tau$
directly to get a convergent series. We could not prove the opposite
either. 

Comparing Eq.~(\ref{example1}) with Eq.~(\ref{rtau}) and 
Eq.~(\ref{example2}) with Eq.~(\ref{moments}), the example demonstrates 
that we can obtain a dependence between the reduced $e^+e^-$ moments $r_n$ 
and the $\tau$ decay rate $r_\tau$ without using an expansion in $d_\tau$. 
This implicit dependence is shown in Fig.~1 for different moments.

In contrast to our previous study~\cite{ptest} the $e^+e^-$ moments can be 
computed without having to perform a sophisticated analysis of divergent 
series and estimating the truncation errors. The only question that remains 
is the question of errors in the present approach. The statistical error 
resulting from the uncertainty in the experimental number for $r_\tau$ can 
easily be taken into account, while for the perturbation series itself we 
suggest to take the difference between results for different $\beta$-function 
accuracy as the resulting error. When analyzed along the contour in the 
complex $Q^2$-plane, the worst pattern of convergence for the 
$\beta$-function expansion is given in the vicinity of the Euclidean point 
$Q^2=m_\tau^2$ and reads
\begin{eqnarray}\label{pattern}
  \beta(0.36)&=&-(0.36)^2(1+0.79(0.36)+1.035(0.36)^2+\rho_3(0.36)^3)
  \nonumber\\&=&-(0.36)^2(1+0.284+0.134+0.0467\rho_3).
\end{eqnarray}
To estimate the error of this expansion we have to estimate the value of 
$\rho_3$. This is also necessary to obtain a feeling for the accuracy of 
the perturbative approximation for the $\beta$-function in 
Eq.~(\ref{pattern}). In the $\overline{\rm MS}$ scheme, this coefficient is 
given by the coefficients of the $\beta$-function up to the known 
coefficient $\beta_3$~\cite{Ritbergen} and the coefficients of the Adler's 
function up to the yet unknown five-loop coefficient $k_3$. In the 
$\overline{\rm MS}$ scheme there exist estimates for this quantity. They 
are essentially Pad\'e estimates valid within this particular scheme and 
result in a value of $k_3=2.17$ or $\rho_3=1.36$. Other estimates are based 
on various optimization procedures for the perturbation series. They give 
values close to the Pad\'e estimate~\cite{KatSta}.

Let us add to the above estimates of $\rho_3$ and present our own analysis. 
The first estimate is a Pad\'e approximation for the $\beta$-function 
itself, which gives $\rho_3=\rho_2^2/\rho_1=1.3548$. We obtain a second 
estimate by considering a one parameter subgroup of the renormalization 
group which leaves the $\beta$-function invariant. It is given by
\begin{equation}\label{rginvbeta}
d'(m_\tau^2)=d(e^\gamma m_\tau^2)
  =d_\tau-\gamma d_\tau^2+(\gamma^2-\rho_1\gamma)
  -(\gamma^3-\frac52\rho_1\gamma^2+\rho_2\gamma)d_\tau^4+\ldots
\end{equation}
and thus expresses $\rho_3$ as a function of $\gamma$. Although the overall 
value of $\rho_3$ is scheme independent by definition, one introduces a 
scheme dependence for $\rho_3$ through the estimation procedure because 
$k_3$ is scheme dependent. The dependence of $\rho_3$ in terms $\gamma$ is 
shown in Fig.~2. We see that the value of $\rho_3=1.36$ in the 
$\overline{\rm MS}$ scheme (i.e.\ for $\gamma=0$) is not stable against 
small variations of $\gamma$. There is, however, a region where $\rho_3$ 
is almost independent of $\gamma$ and yet close to the value given in the 
$\overline{\rm MS}$ scheme. Choosing a scheme in the stability region is 
known as the principle of minimal sensitivity (PMS)~\cite{Ste} and works 
well in a number of applications (see for instance~\cite{opt}). Using this 
principle, we obtain the value $\rho_3=2.4530$ in a wide range around 
the value $\gamma=-1.3288$. It happens that this choice is close to the 
so-called $G$-scheme~\cite{CheKatTka} with $\gamma=-2$, where 
$\rho_3=2.0518$. Although the above dependence is not the most general 
variation within the renormalization group, we believe that it gives an 
additional support for the obtained value of $\rho_3$.

The last estimate may look a bit extravagant, nevertheless the result is 
consistent with that of the previous ones. We fix a scheme such that the 
first few terms of the reduced Adler's function $d(Q^2)$ are represented by 
a pure geometric series
\begin{equation}\label{geom}
d(Q^2)=a_{\rm GS}(Q^2)\Big(1+ka_{\rm GS}(Q^2)+k^2a_{\rm GS}^2(Q^2)
  +k^3a_{\rm GS}^3(Q^2)\ldots\ \Big)
\end{equation}
with coefficients of the $\beta$-function given in the $\overline{\rm MS}$ 
scheme. This is always possible up to order $k^2$ because of the freedom of 
the one-dimensional reparametrization invariance given by 
Eq.~(\ref{rginvbeta}). Numerically one finds
\begin{equation}\label{geomnum}
d(Q^2)=a_{\rm GS}(Q^2)\Big(1-0.1917a_{\rm GS}(Q^2)
  +0.0367a_{\rm GS}^2(Q^2)+(k_3-2.602)a_{\rm GS}^3(Q^2)+\ldots\ \Big)
\end{equation}
where $k_3$ is the unknown five-loop coefficient given in the 
$\overline{\rm MS}$ scheme. By demanding a geometric series behaviour for 
Eq.~(\ref{geomnum}) one calculates $k_3=2.595$. As an average of our three 
estimates we quote $\rho_3=2.0\pm 0.5$. With this final estimate for 
$\rho_3$ we obtain $d_\tau=0.3667\pm 0.0120$ and the results given in the 
last columns of Table~\ref{tab1} and Table~\ref{tab2}.

The statistical error due to the input uncertainties in $r_\tau$ is about
$3\%$ for all moments. This error is larger than the change resulting from 
adding the next (estimated) term of the $\beta$-function expansion. We 
therefore conclude that up to this order in the  $\beta$-function the 
perturbative expansion for the $\beta$-function does not seem to limit the 
accuracy of the resummed predictions for the $e^+e^-$ moments. The main
uncertainty comes from the experimental error in the semileptonic $\tau$ 
decay rate.

To conclude, we have obtained numerical predictions for the moments of the 
$e^+e^-$ annihilation rate in terms of the known $\tau$ decay rate. We 
exploited the analyticity properties of the two-point current-current 
correlator to perform a partial resummation of the perturbative series 
relating the two sets of observables. Our predictions for the $e^+e^-$ 
moments are renormalization scheme independent but depend on the order of 
the perturbative $\beta$-function expansion. We have attempted to estimate 
the error in our prediction for the $e^+e^-$ moments by estimating an 
unknown five-loop piece in the $\beta$-function expansion. Using this 
estimate, we found that the error in our predictions is dominated by the 
experimental error of the $\tau$ decay rate.\\[1truecm]
\noindent{\bf Acknowledgements:}
This work is partially supported by the BMBF, FRG, under contract 
No.~06MZ566. A.A.P. greatfully acknowledges partial financial supported by 
the Russian Fund for Basic Research under contract No. 96-01-01860 and 
97-02-17065. The work of S.G. is supported by the DFG, FRG.

\vspace{1cm}

\centerline{\Large\bf Figure Captions}
\vspace{.5cm}
\newcounter{fig}
\begin{list}{\bf\rm Fig.\ \arabic{fig}:}{\usecounter{fig}
\labelwidth1.6cm\leftmargin2.5cm\labelsep.4cm\itemsep0ex plus.2ex}

\item Implicit dependence of the reduced $e^+e^-$ moments $r_0$, 
$r_1$, $r_2$ and $r_3$ on the semileptonic $\tau$ decay rate

\item Dependence of the $\beta$-function coefficient $\rho_3$ on the 
subgroup parameter $\gamma$ which specifies the choice of the 
renormalization scheme

\end{list}


\begin{thebibliography}{99}
\bibitem{altrev}G.~Altarelli, ``QCD at colliders'',
  CERN preprint No.~CERN-TH/95-196;\\
  ``Status of precision tests of the standard model'',\\
  CERN preprint No.~CERN-TH/96-265, hep-ph/9611239
\bibitem{Shifman}M.~Shifman, Int.~J.~Mod.~Phys.\ {\bf A11} (1996) 3195
\bibitem{Ritbergen}T.~van Ritbergen, J.A.M.~Vermaseren and S.A.~Larin,\\
  ``The four-loop $\beta$-function in quantum chromodynamics'',\\
  Michigan preprint No.~UM-TH-97-01, hep-ph/9701390
\bibitem{mass4}K.G.~Chetyrkin, ``Quark mass anomalous dimension to 
  $O(\alpha_s^4)$'',\\
  Max Planck Inst.~preprint No.~MPI-PHT-97-019, hep-ph/9703278;\\ 
  J.A.M.~Vermaseren, S.A.~Larin and T.~van Ritbergen,\\
  ``The four-loop quark mass anomalous dimension and the invariant quark 
  mass'',\\
  Michigan preprint No.~UM-TH-97-03, hep-ph/9703284 
\bibitem{Pade}J.~Ellis, M.~Karliner and M.A.~Samuel, ``A prediction for the 
  four-loop $\beta$-function in QCD'', CERN preprint No.~CERN-TH/96-327,
  hep-ph/9612202
\bibitem{KatSta}A.L.~Kataev and V.V.~Starshenko,
  Mod.~Phys.~Lett.\ {\bf A10} (1995) 235
\bibitem{AkhZak}R.~Akhoury and V.I.~Zakharov, ``A quick look at 
  renormalons'', hep-ph/9610492.
\bibitem{AltNasRidRen}G.~Altarelli, P.~Nason and G.~Ridolfi,
  Z.~Phys.\ {\bf C68} (1995) 257
\bibitem{Khoze}Y.L.~Dokshitser, V.A.~Khoze and S.I.~Troian,
  Phys.~Rev.\ {\bf D53} (1996) 89
\bibitem{KraPivRen}N.V.~Krasnikov and A.A.~Pivovarov,
  Mod.~Phys.~Lett.\ {\bf A11} (1996) 835;\\
  ``Infrared modified analysis for the $\tau$ decay width'',\\
  Moscow preprint No.~INR-925-96, hep-ph/9607247
\bibitem{PenPivRen}A.A.~Penin and A.A.~Pivovarov,
  ``Numerical analysis of renormalon technique in quantum mechanics'',
  KEK-PREPRINT-96-144, hep-ph/9612204;\\
  ``Practical techniques of QCD vs.\ exact results of solvable models'',\\
  Moscow preprint No.~INR-937-96, hep-ph/9612489
\bibitem{aleph}M.~Davier, ``$\tau$ decays into strange particles and QCD'',\\
  Orsay preprint No.~LAL/96-94;\\
  A.~H\"ocker, ``Vector and axial-vector sprctral functions and QCD'',\\
  Orsay preprint No.~LAL/96-95
\bibitem{ptest}S.~Groote, J.G.~K\"orner, A.A.~Pivovarov and K.~Schilcher,\\
  ``New high order relations between physical observables in 
  perturbative QCD'',\\Mainz preprint No.~MZ-TH/97-09, hep-ph/9703208
\bibitem{resum}S.~Groote, J.G.~K\"orner and A.A.~Pivovarov,\\
  ``Resummation analysis of the $\tau$ decay width using the four-loop
  $\beta$-function'',\\Mainz preprint No.~MZ-TH/97-03, hep-ph/9703268 
\bibitem{tau}A.A.~Pivovarov, Sov.~J.~Nucl.~Phys.\ {\bf 54} (1991) 676;\\
  Z.~Phys.\ {\bf C53} (1992) 461; Nuovo~Cim.\ {\bf 105A} (1992) 813
\bibitem{Braaten}E.~Braaten, Phys.~Rev.~Lett.\ {\bf 60} (1988) 1606;\\
  S.~Narison and A.~Pich, Phys.~Lett.\ {\bf 211 B} (1988) 183;\\
  E.~Braaten, Phys.~Rev.\ {\bf D39} (1989) 1458;\\
  E.~Braaten, S.~Narison and A.~Pich, Nucl.~Phys.\ {\bf B373} (1992) 581
\bibitem{Gru}G.~Grunberg, Phys.~Lett.\ {\bf 95 B} (1980) 70;
  Phys.~Rev.\ {\bf D29} (1984) 2315
\bibitem{Max}C.J.~Maxwell, D.G.~Tonge, ``RS-invariant all-orders 
  renormalon resummations for some QCD observables'', Durham preprint 
  No.~DTP-96-52, hep-ph/9606392
\bibitem{ChetKuhn}K.G.~Chetyrkin, J.H.~K\"uhn and A.~Kwiatkowski, 
  Phys.~Rep.\ {\bf 277} (1996) 189
\bibitem{Pich2}A.~Pich, ``QCD tests from tau decays'', invited talk given 
  at the 20th John Hopkins Workshop ``Non Perturbative Particle Theory and 
  Experimental Tests'' in Heidelberg, June 27--29, 1996, Valencia preprint 
  No.~FTUV/97-03, hep-ph/9701305
\bibitem{Ste}P.M.~Stevenson, Phys.~Rev.\ {\bf D23} (1981) 2916
\bibitem{opt}A.A.~Penin and A.A.~Pivovarov, 
  Phys.~Lett.\ {\bf 367 B} (1996) 342
\bibitem{CheKatTka}K.G.~Chetyrkin, A.L.~Kataev and F.V.~Tkachev,
  Nucl.~Phys.\ {\bf B174} (1980) 345
\end{thebibliography}
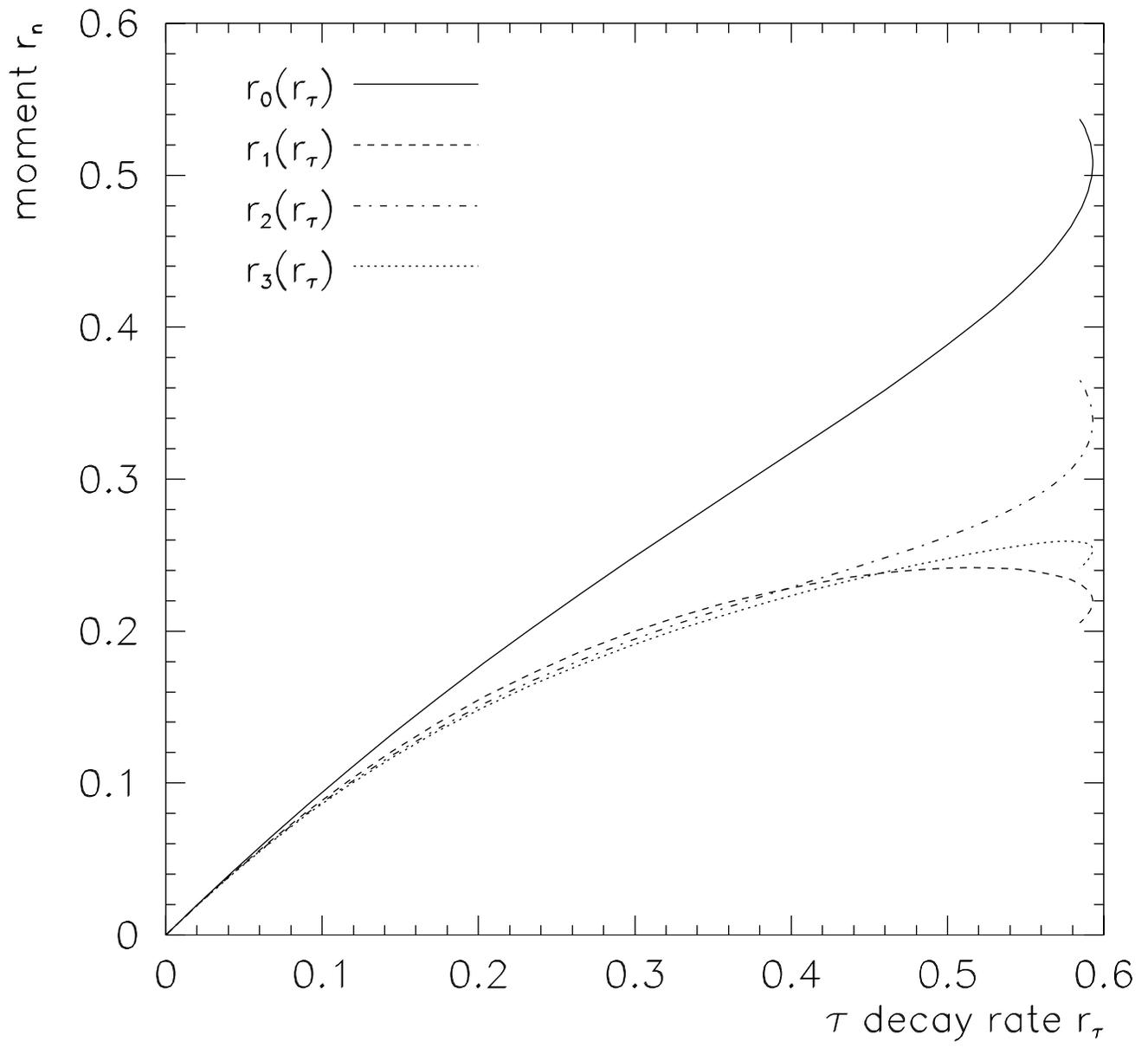
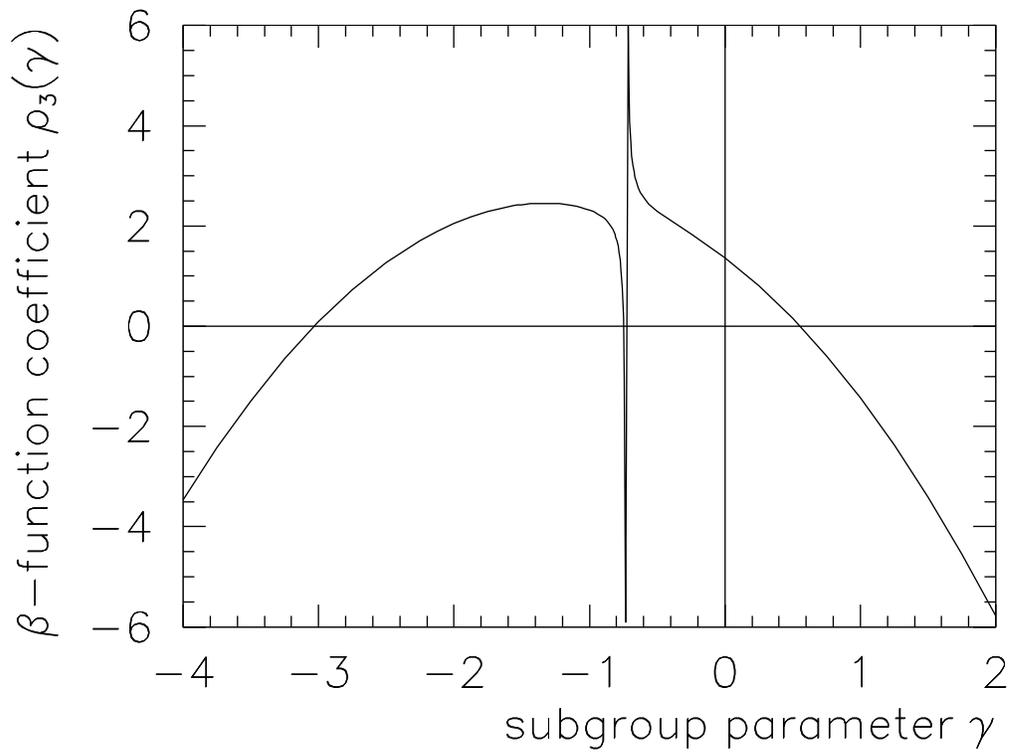
\end{document}